\documentclass[preprint]{aastex}
\newcommand{\HI}{\ion{H}{1}}		% symbol for HI
\newcommand{\HII}{\ion{H}{2}}		% symbol for HII
\newcommand\etal{{\em et~al.}}		% et al., in italics
\newcommand\uv{{\em u,v}}		% u,v, in italics
\newcommand\eg{{\it e.g.}}		% e.g., in italics
\newcommand{\zwa}{{\rm II\,Zw\,70}}	% II Zw 70
\newcommand{\zwb}{{\rm II\,Zw\,71}}	% II Zw 71
\newcommand{\zwab}{{\rm II\,Zw\,70/71}}	% II Zw 70/71
\newcommand\Halpha{{H$\alpha$}}
\newcommand\mjyb{mJy~beam${^{-1}}$}

\newcommand\kms{km~s${^{-1}}$}
\newcommand\Msun{M${_\odot}$}
\newcommand{\Mhi}{$\rm M_{HI}$}
\newcommand\Ho{H$_0$}
\renewcommand\deg{{$^\circ$}}
\newcommand\hour{{$^h$}}
\renewcommand\min{{$^m$}}
\renewcommand\sec{{$^s$}}

\newcommand\gtsim{\lower.5ex\hbox{$\buildrel > \over \sim$}}
\newcommand\ltsim{\lower.5ex\hbox{$\buildrel < \over \sim$}}

\shortauthors{Cox et al.}
\shorttitle{Galaxy Pair II\,Zw\,70/71}

\begin{document}

\title{Stars \&\ Gas in the Galaxy Pair II\,Zw\,70/71}

\author{A.~L.~Cox}
\affil{Department of Physics \&\ Astronomy, Beloit College}
\affil{700 College Street, Beloit, WI  53511-5595}
\email{coxa@beloit.edu}

\author{L.~S.~Sparke}
\affil{Astronomy Department, University of Wisconsin -- Madison}
\affil{475 N. Charter St., Madison, WI 53706}
\email{sparke@astro.wisc.edu}

\author{A.~M.~Watson}
\affil{Instituto de Astronom{\'\i}a, Universidad Nacional Aut{\'o}noma de M\'exico}
\affil{Apartado Postal 72-3 (Xangari), 58089 Morelia, Michoac\'an, M\'exico}
\email{a.watson@astrosmo.unam.mx}

\author{G.~van\,Moorsel}
\affil{National Radio Astronomy Observatory, Array Operations
Center}
\affil{P.O. Box 0, Socorro, NM 87801}
\email{gvanmoor@nrao.edu}

%%%%% ABSTRACT %%%%%%%%%%%%%%%%%%%%%%%%%%%%%%%%%%%%%%%%%%%%%%%%%%%%%%%%%%%%%%%%

\begin{abstract}

\zwb\ (UGC~9562) was classified as a ``probable'' polar-ring galaxy in
the Polar-Ring Catalog (\cite{PRC}), based upon its optical appearance.  
We present 21\,cm and optical observations of this galaxy and 
its companion, the blue star-forming dwarf \zwa.
Our 21\,cm observations show that %0.27 billion 
$5 \times 10^8$ solar masses of
\HI\ is present in a polar ring orbiting 
\zwb, and show a spatially and kinematically contiguous streamer with
%0.14 billion 
$2.5 \times 10^8$ solar masses of \HI\
gas between the two galaxies.  This gaseous bridge, plus
our observations of \Halpha\ line emission in the polar ring and in 
\zwb, are strong evidence for an ongoing interaction between the two
galaxies.  However, the configuration of the streamer 
suggests that the polar ring itself may well have
predated the current interaction, which then stimulated an outburst of
star formation in the ring gas.

\end{abstract}


%%%%% INTRODUCTION %%%%%%%%%%%%%%%%%%%%%%%%%%%%%%%%%%%%%%%%%%%%%%%%%%%%%%%%%%%%

\section{Introduction}
\label{intro}

Polar-ring galaxies (PRG's) are systems with two kinematically distinct
components.  The central component (the ``host galaxy'') is usually a
small S0 galaxy, or occasionally an elliptical galaxy, and the outer
component (the ``polar ring'') is a ring or annulus of material in
a nearly circumpolar orbit about the host.  The fact that the angular
momentum vectors of the host galaxy and of the polar ring are nearly
orthogonal indicates that PRG's are most likely the end products of
mergers or accretion events, rather than a single coherent formation
process.  The Polar-Ring Catalog (\cite{PRC}, hereafter PRC) classifies
6 galaxies as kinematically-confirmed polar rings (category ``A'' in
the PRC), and 27 galaxies as ``probable'' polar rings based upon their
optical appearance (category ``B'' in the PRC).  The catalog is frozen,
but some category ``B'' galaxies have since been confirmed since its
publication (see \cite{SC00}).

The origin of the material in polar rings is something of a mystery,
because the gas masses involved are so large.  Polar rings are
gas-rich, with typically a few billion solar masses of \HI\ gas
(\cite{vGSK87}; \cite{RSS94}; \cite{CS96}; \cite{A97}), and often 
significant molecular gas and dust as well (see \cite{GSS97}).  The
identified polar rings have enough stars that we can see the
rings at optical wavelengths, and stellar rings lie within the rings of
neutral gas (\eg\, \cite{vGSK87}; \cite{A97}).  Because accreted gas
can dissipate energy while the stars cannot, gas and stars will tend 
to settle into very different final configurations.  
The stars now in the polar ring must 
%not have been present in the accreted material, but rather 
have been formed from the gas after it had settled into the thin ring
structure (\eg, \cite{DC94}). 

The host galaxies are small early-type galaxies, typically with
L$_B$~$<$~10$^9$~\Msun.  There is no evidence that the host galaxies
themselves contain substantially more cool gas and dust than other
early-type galaxies without polar rings (see \cite{RSS94}).  Their radio
continuum emission is not significantly stronger than
early-type galaxies without polar rings (\cite{CS94}; \cite{Cetal00}).
This lack of radio continuum emission indicates that any strong
starburst activity which may have been induced by the event that formed
the ring has since ceased.  Even making no allowance for the mass of
stars which have formed in the polar rings, these galaxies have
captured the equivalent of the entire gas content of a typical spiral
galaxy (\eg, \cite{RH94}, hereafter RH94).

It is not clear that a large spiral galaxy could accrete onto a small S0
galaxy without leaving behind a very disorganized system.  If the spiral
companion was destroyed, its stars would not dissipate kinetic energy to
form a ring as the gas has done, but should be visible as faint shells or
diffuse light around the merged system.  Although some PRG's are known to
have shells, many PRG's do not have visible shells (PRC).  If the ring 
material was taken from the outer parts of a spiral galaxy, then we 
should be able to see the remains of this companion, but many polar rings 
have no nearby companions.  If the polar ring was formed by accretion
of a gas-rich dwarf galaxy, there might be no conspicuous stellar shells,
and no visible remnant of the encounter.  But dwarf irregular galaxies at
the present day contain too little neutral gas to make polar rings; the
upper end of the \HI\ mass function for present-day dwarf galaxies is only
a few times 10$^8$~\Msun\ (\eg\ \cite{BR93}; \cite{Br97};
\cite{MvDG98a}).  Searches for clouds of primordial intergalactic \HI\
indicate that there is very little gas in this form at the current
epoch (\eg\ \cite{Br93}; \cite{Tetal96}; \cite{Zetal97}).  If PRG's are
stable structures, most of them may have formed early in the history of
the universe, when large clouds of intergalactic gas should have been
more common.

In any case, it is highly probable that the accretion of material onto
many host galaxies is not complete.  The gas of the polar ring has come
from an orbit of high angular momentum with a large impact parameter, and
it is likely that streamers of gas would remain around most newly-formed
polar rings well after the original encounter (\cite{HM95}).  The study
of probable polar-ring systems which have nearby companions are therefore
very likely to reveal gaseous tails or bridges which may help us better
understand the interactions which form polar rings.

\zwb\ (UGC~9562) is classified as ``B-17'', a ``probable'' polar ring
system, in the PRC.  An optical image of this galaxy and its blue dwarf
companion, \zwa\, is shown in Figure~\ref{rcont}; the polar ring runs
NE to SW.  Optical spectroscopy in the \Halpha\ emission line of
\zwb\ by Reshetnikov \&\ Combes (1994) revealed two gaseous components
with orbits nearly perpendicular to each other --- one with a velocity
gradient along the apparent major axis of the polar ring, and another
with a velocity gradient along the apparent major axis of the host
galaxy.  \zwb\ is thus a kinematically-confirmed polar-ring galaxy.
Earlier observations of this system by Balkowski \etal\ (1978;
hereafter BCW78) in the 21\,cm line with the Westerbork Synthesis Radio
Telescope (WSRT) also showed rotation of gas along the polar ring, and
in addition detected a cloud of \HI\ gas between the two galaxies.  No
optical counterpart brighter than 25.5~mag\,arcsec$^{-2}$ to this cloud
was detected in {\em B} band.  The \zwab\ system is thus one of the
best known candidates for a polar ring in the process of formation by
accretion of gas from a nearby companion.

We have obtained observations in the 21\,cm line of \HI, using the Very
Large Array
(VLA\footnote{The National Radio Astronomy Observatory is a facility
     of the National Science Foundation operated under cooperative
     agreement by Associated Universities, Inc.}) 
aperture synthesis telescope.  These observations are more sensitive
than the observations by BCW78, and have approximately twice the spatial
and velocity resolution.  We have also obtained \Halpha\ and broad-band
optical images of this system, to search for evidence of ongoing star
formation and for an optical counterpart of the gas cloud between
\zwa\ and \zwb.  The \HI\ distribution and kinematics of the \zwab\
system are presented in Section~\ref{21cm}, and the optical data in
Section~\ref{optical}.  In Section~\ref{discussion}, we discuss the
morphology of the stars and gas in this system, and investigate the
possibility that the polar ring around \zwb\ is the product of an ongoing
accretion event from gas originating in \zwa.  We take 
\Ho~=~75~km~s$^{-1}$~Mpc$^{-1}$, implying a distance of 18.1~Mpc for the
system.

%%%%% SECTION %%%%%%%%%%%%%%%%%%%%%%%%%%%%%%%%%%%%%%%%%%%%%%%%%%%%%%%%%%%%%%%%%

\section{21\,cm Observations}
\label{21cm}

The \zwab\ system was observed with the ``C'' and ``D'' configurations
of the VLA.  The C-configuration data were obtained on 02 December
1994, and the D-configuration data on 02 May 1995.  The instrumental
parameters for the observations are summarized in Table~\ref{instparms}.
Our velocity resolution was $\simeq$\,2.6~\kms.  Because of the high
velocity resolution, we observed the \zwab\ system with two independent
bandpasses (IF's), centered at 1232 \&\ 1268~\kms.  This observing mode
allowed us to have a wide enough total bandwidth to include all of the
gas in both systems, without sacrificing velocity resolution.

% TABLE 1
\placetable{instparms}
% Table with instrumental parameters for observations

Calibration and mapping were performed using the NRAO Astronomical
Image Processing System (AIPS).  The data from each configuration
and each IF were calibrated and continuum-subtracted independently.
Standard calibration procedures were followed, as described in Appendix
B of the {\em AIPS Cookbook} (\cite{Cox94}).  As the IF's were separated
in velocity space, each IF contained channels which were free of line
emission on only one side of the bandpass.  A continuum dataset was
constructed for each IF by averaging together the line-free channels
in the \uv\ plane; this dataset was subtracted from all channels in
that IF to give one continuum-subtracted set of line data for each IF.
This method of continuum subtraction assumes that the continuum spectrum
is flat over the bandpass of the observation, and that there are no strong
continuum sources.  Both of these criteria were satisfied by this dataset.
Faint 21\,cm continuum emission was detected at the positions of \zwa\
and \zwb; a continuum map is shown in Figure~\ref{rcont}, and the nature
of this emission is discussed further in Section~\ref{continuum}.

%FIGURE 1
\placefigure{rcont}
% Radio continuum contours for II Zw 70/71, overlaid on B band image

After continuum subtraction, the end channels were discarded and all four
datasets (two IF's for each configuration) were combined in the \uv\
plane to yield a single, continuum-subtracted dataset for the \zwab\
system.  Maps were made and corrected for instrumental response using
the robust imaging task, IMAGR.  Robust weighting improves the shape
of the synthesized beam dramatically for multi-configuration datasets,
as these data have even higher concentrations of short baselines than
single-configuration data (\cite{Br95}; see also Section~5.2.3 of the
{\em AIPS Cookbook}).  The resulting channel maps have a spatial 
resolution of 22.5\arcsec\,$\times$\,20\arcsec, and include structures
with sizes up to 10\arcmin; they are shown in Figure~\ref{chanmaps}.

% FIGURE 2
\placefigure{chanmaps}
% Figure showing channel maps corrected for instrumental response

%%%%% SUBSECTION %%%%%

\subsection{Continuum Emission}
\label{continuum}

As shown in Figure~\ref{rcont}, faint 21\,cm continuum emission was
detected at the positions of \zwb\ and its companion galaxy, \zwa.
The emission observed in \zwb\ is centered on an \HII\ region in the
polar ring; no emission is observed in the central portion of the
host galaxy.  Because the continuum map is constructed from an average
of the line-free channels, the RMS noise in the continuum map is only
0.2~\mjyb, significantly lower than the RMS in the individual line
channels (see Table~\ref{instparms}).  As part of a radio continuum
survey of polar-ring galaxies (\cite{CS94}; \cite{Cetal00}), we have
also obtained 6cm continuum observations of these sources.  The 6cm
observations were made in the VLA ``B'' configuration, which gives a
synthesized beam of approximately 5\arcsec.  The on-source integration
time was only 30 minutes, but because of the larger bandwidth of the
continuum observations (50~MHz), the RMS noise in the 6cm maps is
0.09~\mjyb.  The 6cm emission which was detected was also unresolved,
and at the same positions as the 21\,cm emission.

As the sources are unresolved at both wavelengths, differences in baseline
coverage between the two sets of observations should not present a problem
when calculating radio spectral indices.  In both \zwa\ and \zwb, the 
spectral indices (defined by S~$\propto\,\nu^\alpha$) are quite steep, with
$\alpha\,\simeq\,-1$.  This is consistent with radio spectral indices
of star-forming galaxies (\eg\, \cite{Cetal91}).  Faint, steep-spectrum
radio continuum emission is not uncommon in polar-ring galaxies
(\cite{Cox96}), and the emission may be either centrally concentrated
or extended along the polar ring (\cite{A97}; \cite{CvG92}).

%%%%% SUBSECTION %%%%%

\subsection{Moment Maps}
\label{momentmaps}

Using the cleaned map cube at 5.2~\kms\ velocity resolution, we
constructed moment maps of the velocity profile at each point.  To
improve the sensitivity to low-level extended emission, we first
smoothed the entire map cube to 40\arcsec\ resolution.  We then
selected the areas which contained real line emission in this smoothed
cube, using a two-step process.  First, we applied a uniform flux
cutoff to the entire cube, eliminating all pixels with an intensity
less than 1~\mjyb\ (twice the RMS noise in the smoothed maps).  To
minimize the confusion from noise peaks not related to real line
emission, we then inspected each velocity channel visually, and
selected only those areas which appeared to contain line emission,
and blanked all other areas.  These smoothed and
clipped channel maps were then used as masks; we calculated moment maps
from the cube at full resolution, including only the flux from regions
where the smoothed maps were found to have emission, and the flux in
the full-resolution maps was above a 2-$\sigma$ cutoff level of
0.8~\mjyb.

%FIGURE 3
\placefigure{mom0}
% total-intensity map of II Zw 70/71 system and pos-vel plots

%FIGURE 4
\placefigure{mom1}
% velocity field for II Zw 70/71 system, overlaid on total intensity map

The total-intensity map at 21\,cm (Figure~\ref{mom0}) shows significant
quantities of neutral gas in the PRG (\zwb), its dwarf companion (\zwa)
about 260\arcsec\ away, a streamer which appears to connect the two,
and an extension of the gas to the east of the polar ring.  The
polar-ring gas extends about 50\arcsec\ to the north and 70\arcsec\ to
the south, and looks ``C'' shaped; the ring seems to bend towards
\zwa\ at both ends.
  
Figure~\ref{mom0}b is a position-velocity cut taken along the
optical major axis of the polar ring.  In general, the rotation of the gas
is along the ring major axis, but there is a larger spread
in velocity and space on the southern side of the ring than on the
northern side.  Figure~\ref{mom0}c is plotted along the position angle
of the \HI\ streamer seen between the two galaxies.  To increase our
signal-to-noise for the faint streamer, the data for this
position-velocity profile were summed out to a distance of 1\farcm5
perpendicular to this axis, on each side of the streamer.  This
includes all 21\,cm line emission detected in these observations.  
The streamer is contiguous with both galaxies, in both position and
velocity space --- a strong indication of an ongoing interaction
between \zwa\ and \zwb.   However, the gas that extends to the east of
the polar ring appears to be a continuation of the streamer in
both position and velocity, which suggests that the streamer passes 
in front of the polar ring, or behind it. 

The intensity-weighted velocity field for the \zwab\ system is shown in
Figure~\ref{mom1}.  The optical
appearance of the polar ring in \zwb\ suggests that we see it almost
edge-on.  With a rotation speed of 95\kms, 
%and Ho = 75
this gives a dynamical mass
of 19 billion \Msun\ within a 50\arcsec\ distance from the center of
the galaxy, or M/L$_B \approx 21$.  This is much larger than the value 
M/L$_B = 2.8$ within 30\arcsec, derived by \cite{RC94} from long-slit
spectroscopy of the ring in H$\alpha$.  The position-velocity cut of
Figure~\ref{mom0}b shows no sign that the rotation speed of the ring
gas is dropping off, even far outside the optical radius of the
host galaxy in \zwb\ (18\arcsec; see \cite{Sa91}).  Thus a significant
amount of the mass in this galaxy must be in the form of dark matter.

The gas of the companion galaxy, \zwa, exhibits
regular rotation about its apparent optical minor axis.  This is in
contrast with the earlier observations by BCW78, where the rotation
appeared to be skewed from the minor axis, so that the kinematic major
axis was oriented almost exactly east-west.  However, their
observations probably confused gas associated with \zwa\ with that from
the streamer of \HI\ between the two galaxies.  This problem is reduced
in the current observations by our improved spatial resolution
($\simeq$~21\arcsec\ for these observations, compared with
$\simeq$~34\arcsec\ for the observations of BCW78).  
If \zwa\ is also seen close to edge-on, our observed rotation speed of
67\kms\ within a 45\arcsec\ radius gives a dynamical mass of
%Ho = 75
8.3 billion \Msun\ for this galaxy.

%%%%% SUBSECTION %%%%%

\subsection{Line Profiles}

% TABLE 2
\placetable{physq}
% Table with physical quantities derived from HI observations

A 21\,cm line profile for this system was derived by summing the flux
for each channel over the area containing line emission in the 21\,cm
total-intensity map of Figure~\ref{mom0}.  The 21\,cm spectrum for the
entire region is shown as a solid line in Figure~\ref{spectrum}.  The
total \HI\ gas mass we derive for this system is
%Ho = 75
$ 9 \times 10^8$~\Msun, consistent with the results from BCW78
and \cite{RSS94}.  The 21\,cm spectrum is very asymmetric, without the
typical double-horned shape typical of spiral galaxies and other PRG's.
This is not particularly surprising, however, as gas from the entire
system is included, and the gas in the ring is not symmetric about
the galaxy center.

As can be seen in Figure~\ref{mom0}, it was impossible to distinguish
the emission from \zwb\ from that of the \HI\ streamer in all of the
velocity channels.  We constructed approximate line profiles for each
of the components by selecting regions for \zwa\ and \zwb\ in the
total-intensity map and assuming that all emission in these regions
belongs to these galaxies, and not to the streamer.  The resulting
global profiles for each component are shown as dashed lines in the
figure.  Because of the method we used in determining these profiles,
the amount of gas belonging to \zwb\ in the 1250--1300~\kms\ velocity
range is probably an overestimate, while in the same velocity range we
underestimate the amount of gas in the \HI\ streamer.  Based upon these
global profiles, we estimate that 20\%\ of the neutral hydrogen is
associated with \zwa, 25\%\ with the streamer, and the remaining
55\%\ with the polar ring around \zwb.  Based upon the use of a
variety of different boxes for these regions, we estimate that the
uncertainty introduced by our separation methods to be 10\%.

For the dwarf galaxy \zwa, we derive M$_{\rm HI}$/L$_B$\,$\simeq$\,0.3.  
Though substantial, the \HI\ mass of \zwb\ is one of the smallest 
gas masses associated with known PRGs  (\eg\, \cite{RSS94}; see
Section~\ref{intro}).  However, it is also rather small, 
and faint at optical wavelengths; so the ratio 
M$_{\rm HI}$/L$_B$\,$\simeq$\,0.55 is only slightly below the median 
for confirmed polar rings (\cite{SC00}).  
These galaxies are both rather gas-rich, with values of 
M$_{\rm HI}$/L$_B$ typical of late-type spirals or gas-rich dwarf 
irregular galaxies (RH94; see Table~\ref{physq}).  
For the polar ring considered alone, 
M$_{\rm HI}$/L$_B$\,$\simeq$\,2.4, which is more typical of the blue
low-surface-brightness disk galaxies in the sample of \cite{MG97}: see
Figures~6 and 7.

%FIGURE 5
\placefigure{spectrum}
% Global 21cm profiles for the II Zw 70/71 system

%%%%% SECTION %%%%%%%%%%%%%%%%%%%%%%%%%%%%%%%%%%%%%%%%%%%%%%%%%%%%%%%%%%%%%%%%%

\section{Optical Observations}
\label{optical}

Optical images of the \zwab\ system were obtained with the 1.8-meter
Perkins telescope of the Ohio Wesleyan University and the Ohio State
University at Lowell Observatory. We used the Ohio State
University Imaging Fabry-Perot Spectrometer, without an etalon, as a
focal-reducing camera.  Observations were conducted on the night of
1995 April 8 through variable cirrus, and we obtained images totaling
5400 s in $B$, 1800 s in $V$, 2100 s in $R$, and 5400 s in a
narrow-band H$\alpha$ filter. We observed again on the night of 1995
June 5 under photometric conditions and obtained shorter exposures in
the same filters. The June images were calibrated using standards from
Landolt (1983) and Barnes \& Hayes (1982) and this calibration was
transfered to our deeper April data using field stars.
We produced a continuum-free H$\alpha$ image by scaling and subtracting
the $R$ image.  Table~\ref{mags} gives our derived magnitudes and colors.

% FIGURE 6
\placefigure{BVimages}
% optical images of II Zw 71/70 system

% TABLE 3
\placetable{mags}
% Optical magnitudes & colors for the system

In the $B$ band image, and the $B-R$ color map of Figure~\ref{BVimages},
% we don't show the B image!
the polar ring appears very bumpy, with bright knots along it.
There is also strong bending of the ring component, beginning
fairly close to the central galaxy.  The ring is also asymmetric, with
the northern side brighter than the southern side.  The close
companion, \zwa, has irregular plumes extending from the main body at
optical wavelengths.  Our exposures were not deep enough to improve
on the limit derived by BCW78 for an optical counterpart to the gas
streamer.

The global colors of both galaxies are similar to those of late-type
spiral galaxies (\eg, RH94).  The blue color of \zwb, however, is
partly due to the light from the polar ring.  We averaged in boxes over
representative portions of the polar ring and the host galaxy (see
Figure~\ref{BVimages}), and found $B-R \simeq 0.99$ for the host 
galaxy, while the ring is much bluer, with $B-R \simeq 0.65$ and $B-V
\simeq 0.36$.  The ring is significantly bluer than average 
for the disk of a late-type galaxy in the sample of 
de Jong (1995, 1996).  However, both its color and its gas content are
similar to the gas-rich low-surface-brightness disks of \cite{MG97}; 
the outer disks of the blue low-surface-brightness galaxies of
\cite{Bell00} also have similar optical colors.

The total $V$-band magnitude of \zwb\ ($M_V = - 17.3$) puts it about 
a magnitude below the mean Tully-Fisher relation defined by bright galaxies 
(\cite{PT92}).  However, \zwb\ lies approximately on the Tully-Fisher
relation for gas-rich extreme late-type galaxies (\cite{MvDG98b}),
which have similar ratios of M$_{\rm HI}$/L$_B$.

The continuum-subtracted \Halpha\ image of \zwa\ in
Figure~\ref{BVimages} shows ongoing star formation in the center.
We also see filamentary \Halpha\ emission in the southwestern plume,
but not in the northeastern plume.  Our measured line flux of 
$6.9 \times 10^{-13}~$erg~cm$^{-2}$~s$^{-1}$ yields a total energy 
output of $2.5 \times 10^{40}$~erg~s$^{-1}$ in \Halpha.
Assuming that the initial mass function follows Salpeter's form, we
can use Kennicutt's (1998) recipe to infer that \zwa\ is producing 
0.2\Msun~yr$^{-1}$ of new stars.  At this rate, it would have taken 
3~$\times$~(M/L$_B$)~Gyr to make the galaxy's present population of
stars, where M/L$_B$ is the mass-to-light ratio in solar units.
If the rate stays constant in the future, and we neglect gas recycled
from aging stars, then the observed \HI\ gas will be exhausted in 
0.9~Gyr; this timescale is typical of starbursting irregulars 
(\cite{H97}). 

The \Halpha\ emission of \zwb\ is from bright knots throughout the
polar ring.  The knot near the center could be either an \HII\ region
in the ring that is simply projected close to the center, or within
the central galaxy itself.  The ring emission is 
asymmetric.  Even though the \HI\ gas is more extended to the south,
the northern half of the ring is much stronger in \Halpha,
and is also the region where we detected the radio continuum emission 
shown in Figure 1.  The measured \Halpha\ flux of 
$1.2 \times 10^{-13}~$erg~cm$^{-2}$~s$^{-1}$ is significantly larger
than that estimated by  Reshetnikov \&\ Combes (1994) 
from their slit spectrum.   
It translates to a total of $4.4 \times 10^{39}$~erg~s$^{-1}$ 
in \Halpha, corresponding to only 0.035\Msun~yr$^{-1}$ of new stars.  
Although the polar ring has almost the same colors as \zwa, it is
making its stars much more slowly, at a pace more typical of
irregular galaxies (\cite{H97}).  At this rate, the observed \HI\ 
gas would be consumed only over 14~Gyr, while it would have taken
6~$\times$~(M/L$_B$)~Gyr to make the ring's present population of
stars.  However, Figure~3 of \cite{Bell00}, using models in 
which star formation increases or decreases monotonically in time,
shows that the ring's observed blue color corresponds to a constant, 
or even rising, rate of starbirth.  Thus despite its messy appearance, 
the ring in \zwb\ is likely to be considerably older that the 
$\simeq$~1~Gyr suggested on morphological grounds by 
Reshetnikov \&\ Combes.

%%%%% SECTION %%%%%%%%%%%%%%%%%%%%%%%%%%%%%%%%%%%%%%%%%%%%%%%%%%%%%%%%%%%%%%%%%

\section{Discussion: the Polar-Ring of II\,Zw\,71}
\label{discussion}

% Point 1:  Small but gas-rich; likely lots of DM

We observe \HI\ gas in regular rotation about the minor axis of the
optical polar ring in \zwb, and none that shares the rotation of the S0
disk.  \zwb\ is thus a confirmed polar-ring galaxy.  The amount of
\HI\ is fairly modest, compared with other polar rings (\cite{RSS94}), 
but is much greater than normal for an S0 galaxy, and is more typical
of a late-type spiral or large irregular galaxy (\cite{RH94}).  As with
other polar rings, the extent of the \HI\ gas is greater than that of
the optical ring.  The rotation curve does not drop, even far outside
the optical radius of the host galaxy.  As in other polar-ring galaxies,
it appears that most of the mass in this galaxy is dark (\eg, \cite{SWR83};
\cite{WMS87}; \cite{SRJF94}; \cite{A97}).

% Point 2:  This ring is either new or recently disturbed

The polar ring of \zwb\ is asymmetric, at both optical and radio
wavelengths.  In addition, observations at \Halpha\ show clumpy
emission in the ring.  This emission is more pronounced on the northern
side of the galaxy, the same side on which faint, steep-spectrum radio
continuum emission is observed.  The presence of both \Halpha\ emission
and radio continuum emission along the ring, and its blue color, are
strong arguments for ongoing star formation.  
The asymmetry of the ring material and the evidence of star
formation in the ring make it likely that new material has recently
been added to the polar ring, or that pre-existing ring material has
recently been disturbed by a new interaction. 

% Point 3:  This PR is in a gas-rich environment, and is interacting

At our assumed distance of 18.1~Mpc, 
%(for \Ho~=~$75~km~s^{-1}~Mpc^{-1}$),
the projected separation of 4.4\arcmin\ between these two galaxies
corresponds to a separation of $\simeq$~23~~kpc.  At a relative speed
of 100~\kms, it would have taken these galaxies at least 230~Myr to
separate after a close interaction.  The orbital timescale for gas in
the outer regions of \zwb\ is about 200~Myr, implying that the most
recent interaction between these two galaxies may have taken place
within a few orbital timescales. 
% Reshetnikov \&\ Combes (1994) estimate the age of the polar ring to be
% \ltsim~1~Gyr, based upon its morphology, and the kinematics of the
% ring in the \Halpha\ line.  

Was the polar ring of \zwb\ formed by accretion from the companion 
galaxy \zwa?  The two galaxies are smoothly connected in both position
and velocity by a gaseous bridge or streamer, in which we observe no
stellar component.  Assuming that the bridge is the remnant of the
interaction that created the polar ring, so that all of the gas from
the polar ring and streamer originally belonged to \zwa, would yield
$\rm{M_{HI}/L_{B}}~=~0.55$ prior to the interaction --- a high, but
not unreasonable, value for an irregular galaxy.
Since the dynamical mass of \zwb\ is more than twice that of \zwa, 
the outer layers of a gas-rich system the size of \zwa\ would not 
have been very tightly bound, and might have been fairly easily detached.

% OLD VERSION OF ABOVE PARAGRAPH 
%
% \zwb\ is smoothly connected to its companion galaxy, \zwa, by a gaseous
% streamer.  We observe no stellar component to this bridge.  Assuming that
% all of the gas from the polar ring and streamer originally belonged to
% \zwa\ yields a high, but not unreasonable, value of $\rm{M_{HI}/L_{B}}$
% for an irregular galaxy.  Is this a new polar ring in the process of
% formation by accretion from \zwa?  The projected separation between
% these two galaxies is 4.4 minutes of arc --- at our assumed distance of
% 13.6~$h^{-1}$~Mpc, this is $\simeq$~17.5~$h^{-1}$~kpc.  At a relative
% speed of 100~\kms, it would have taken these galaxies less than 200~Myr
% to separate after a close interaction.  Since the dynamical mass of \zwb\
% is more than twice that of \zwa\, the outer layers of a gas-rich system
% the size of \zwa\ would not be very tightly bound.  The orbital timescale
% for gas in the outer regions of \zwb\ is, also 200~Myr, implying that
% very few dynamical timescales of the ring have passed since the most
% recent interaction between these two galaxies.

% Point 4:  Unlikely that new material formed ring
The blue stars and \Halpha\ regions in the polar ring are spread along
a line, implying that much of the gas has already settled into a disk
or ring, that we now see almost edge-on.  It is difficult to explain
how gas that was accreted only a few orbital timescales ago could have
already settled into a plane.  If there were enough dissipation to
cause flattening on this short timescale, we would expect substantial
amounts of accreted gas to be falling into the center of the host
galaxy, and either forming stars there, or fueling a central black
hole.  However, this does not seem to be the case.  Our observations
reveal no radio continuum source, and 
%no strong \Halpha\ emission at the center of the host galaxy; 
the far-infrared IRAS flux is also low
(PRC; \cite{RSS94}).  The ionized gas of the central galaxy does not
share the rotation axis of the ring material (\cite{RC94}), so it is
unlikely to have come from the ring.

% Point 1:  Small but gas-rich; likely lots of DM
% Point 2:  This ring is either new or recently disturbed
% Point 3:  This PR is in a gas-rich environment, and is interacting
% Point 4:  Unlikely that new material formed ring
% Point 5:  Alternative hypothesis - pre-existing ring, new interaction

The contiguous streamer of gas between \zwa\ and \zwb, as well as the
star formation activity in both \zwa\ and the polar ring, indicate an
ongoing interaction between these two galaxies.  However, the fact
that HI gas to the east of the polar ring appears to be a continuation
of the streamer to the west, in both position and velocity, implies
that the streamer probably lies behind \zwb\ or in front of it, rather
than merging into the polar ring.  That is not what we would expect if
both the polar ring and the streamer originated in the same, recent,
accretion event.
%Also, the \HI\ gas of the streamer
%lies in a plane almost 90\deg\ inclined to the polar ring. 
A further indication that the ring itself has existed for some time
comes from the relatively weak \Halpha\ emission.  At the present rate of 
star formation, it would take several gigayears to build up the stars 
that we now observe in the ring, while the blue color indicates that
starbirth was no more rapid in the past than it is now.
We suggest instead that the polar ring of \zwb\ predated the most
recent close passage of \zwa, but was disturbed by it.  Just as in the
disks of spiral galaxies (Larson \& Tinsley 1978, Mihos \& Hernquist
1996, Barton, Geller \& Kenyon 2000), tidal torques have `rejuvenated'
the ring, causing the vigorous star formation that we now observe.

%However, the observed planarity of the ring material and the regular
%rotation of the ring seems to indicate that the polar ring is older
%than the current interaction.  Perhaps the polar ring in \zwb\
%existed before the current interaction.  It may have been formed
%during an earlier passage of \zwa\ or another source entirely, and is
%now being disturbed by the current interaction.

%%%%% ACKNOWLEDGMENTS %%%%%%%%%%%%%%%%%%%%%%%%%%%%%%%%%%%%%%%%%%%%%%%%%%%%%%%%%

\acknowledgments 
ALC and LSS acknowledge support from the National Science Foundation through
grant AST-9320403.  The work reported here forms part of the PhD
thesis of Andrea Cox, who was an NRAO Predoctoral Fellow while
much of it was carried out.  We would like to thank Michael
Rupen and Jay Gallagher for helpful discussions.

%\singlespace

%%%%% BIBLIOGRAPHY %%%%%%%%%%%%%%%%%%%%%%%%%%%%%%%%%%%%%%%%%%%%%%%%%%%%%%%%%%%%

\newpage
\singlespace

%%%%% DUMMY TABLES (include these to get references right!) %%%%%%%%%%%%%%%%%%%

\begin{table}
\dummytable\label{instparms}
\end{table}

\begin{table}
\dummytable\label{mags}
\end{table}

\begin{table}
\dummytable\label{physq}
\end{table}

\begin{table}				% This is here because AASTeX 
\dummytable\label{notarealtable}	% doesn't seem to want to read in
\end{table}				% the last dummytable.

\clearpage	% this has to be here because AASTeX gets confused

%%%%% FIGURE CAPTIONS (on a new page!) %%%%%%%%%%%%%%%%%%%%%%%%%%%%%%%%%%%%%%%%

%\newpage

%\caption[{\em continued}]{{\em continued}}

\includegraphics[width=\linewidth]{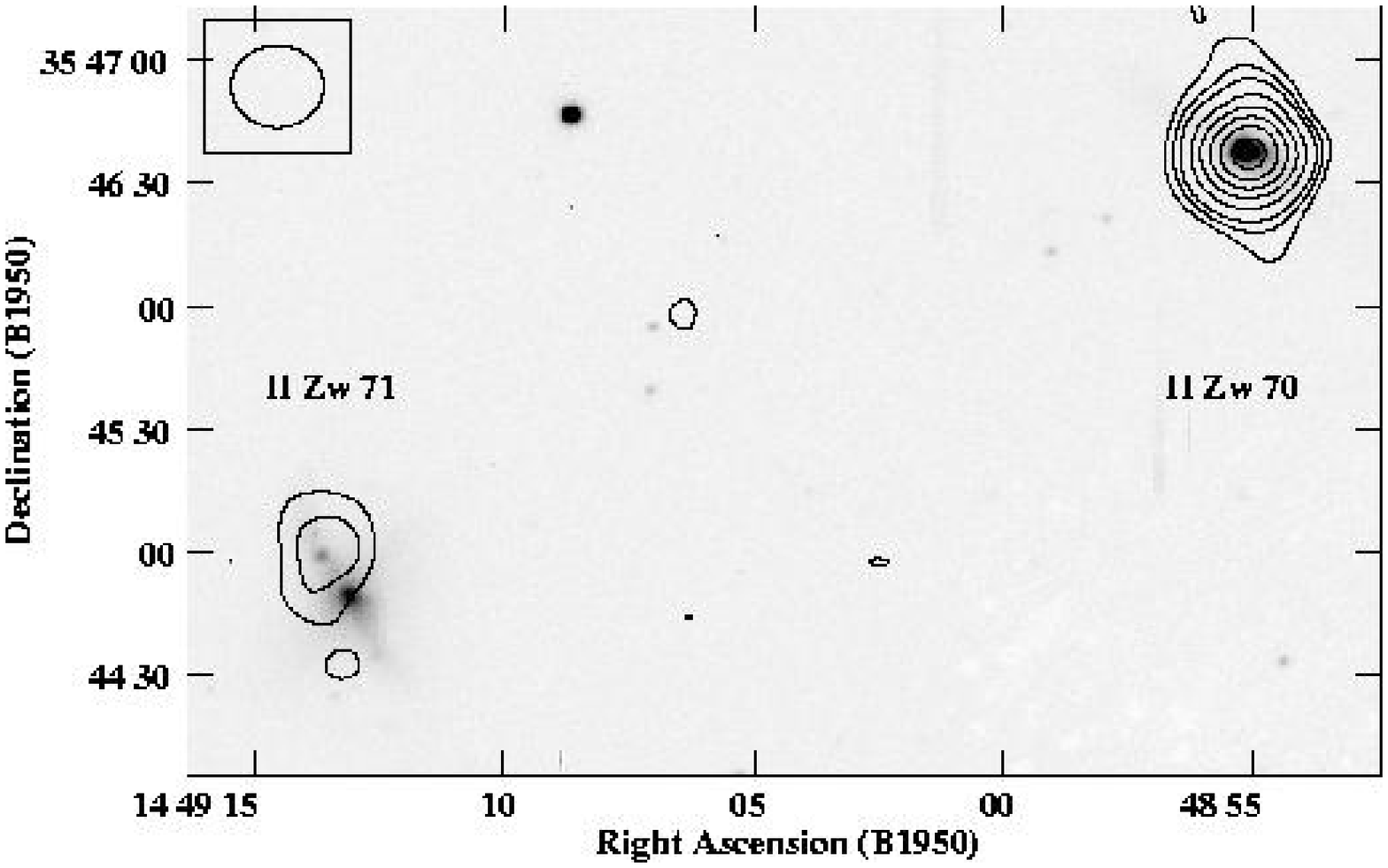}

\figcaption[II~Zw~71: Radio continuum map]{
%%% 21cm CONTINUUM MAP, OVERLAID ON B band IMAGE
  21\,cm radio continuum contours for the II~Zw~71/70 system, overlaid
  on the optical (B band) image taken with the Ohio State University
  Imaging Fabry-Perot Spectrometer (IFPS).  The continuum map was
  constructed from an average of the line-free channels.  II~Zw~71
  (on the left) is
  the candidate polar-ring galaxy, and II~Zw~70 (right) is the dwarf
  companion.  The peak flux is 4.5~mJy~beam$^{-1}$, and the solid
  contour levels are at 3, 4.5, 6, 9, 12, 15, \&\ 18-$\sigma$, where
  $\sigma$\,$=$\,0.2~mJy~beam$^{-1}$.
\label{rcont}}

\clearpage

\includegraphics[width=\linewidth]{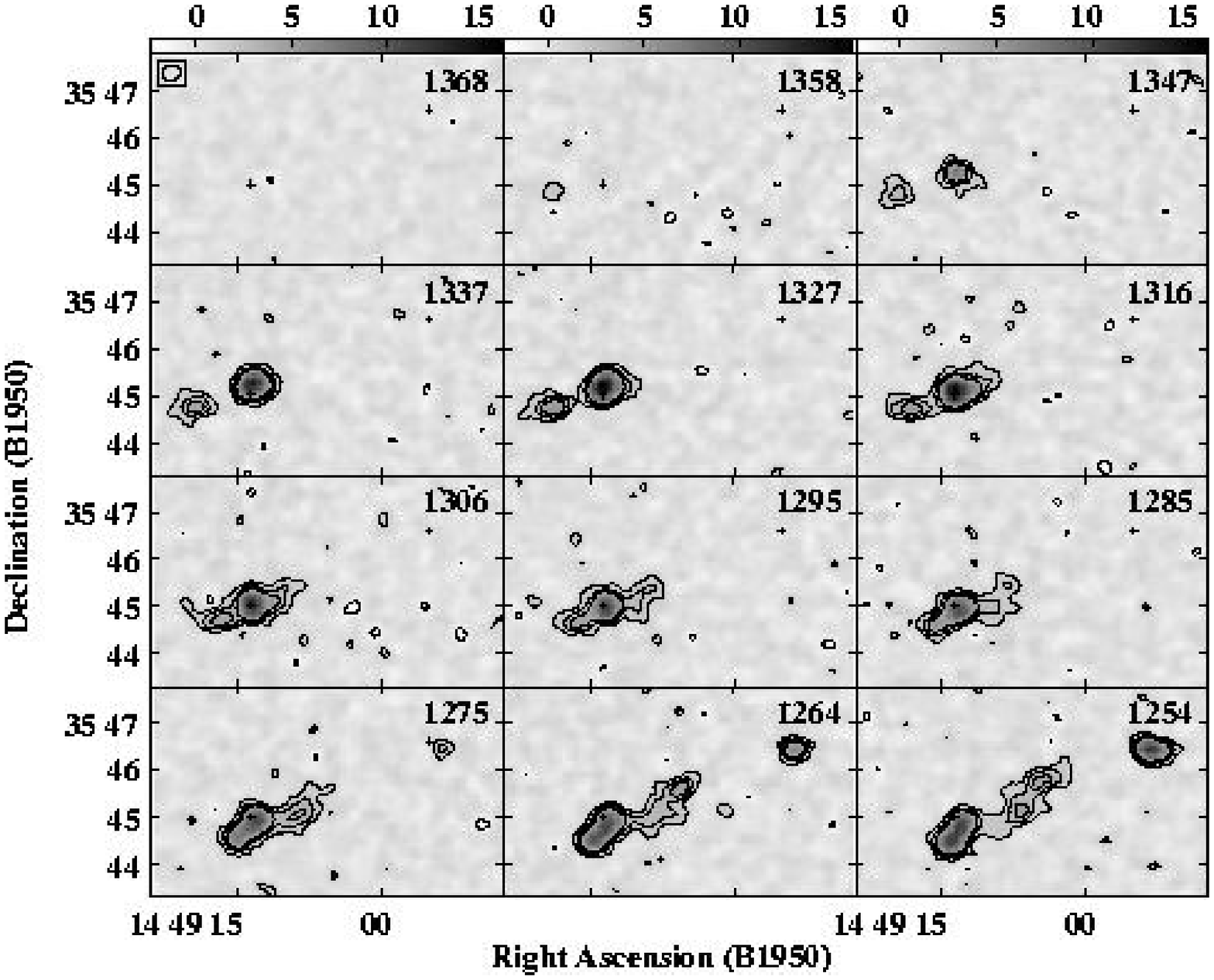}

\figcaption[II~Zw~71: 21\,cm channel maps]{
%%CHANNEL MAPS
  21\,cm channel maps for the II~Zw~71/70 system, smoothed to
  10~\kms\ velocity resolution.  The crosses mark the central
  positions of II~Zw~71 (east) and II~Zw~70 (west).  The synthesized
  beam is shown in the upper left-hand corner of the first channel,
  and each channel is labelled with its central velocity
  in \kms.  Dashed contours are at $-$3-$\sigma$; solid contours are
  at 3, 6, \&\ 9-$\sigma$ ($\sigma\,=\,0.4$~mJy~beam$^{-1}$).
  The greyscale levels are shown (in mJy) at the top of the image.
\label{chanmaps}}

\clearpage

\includegraphics[width=\linewidth]{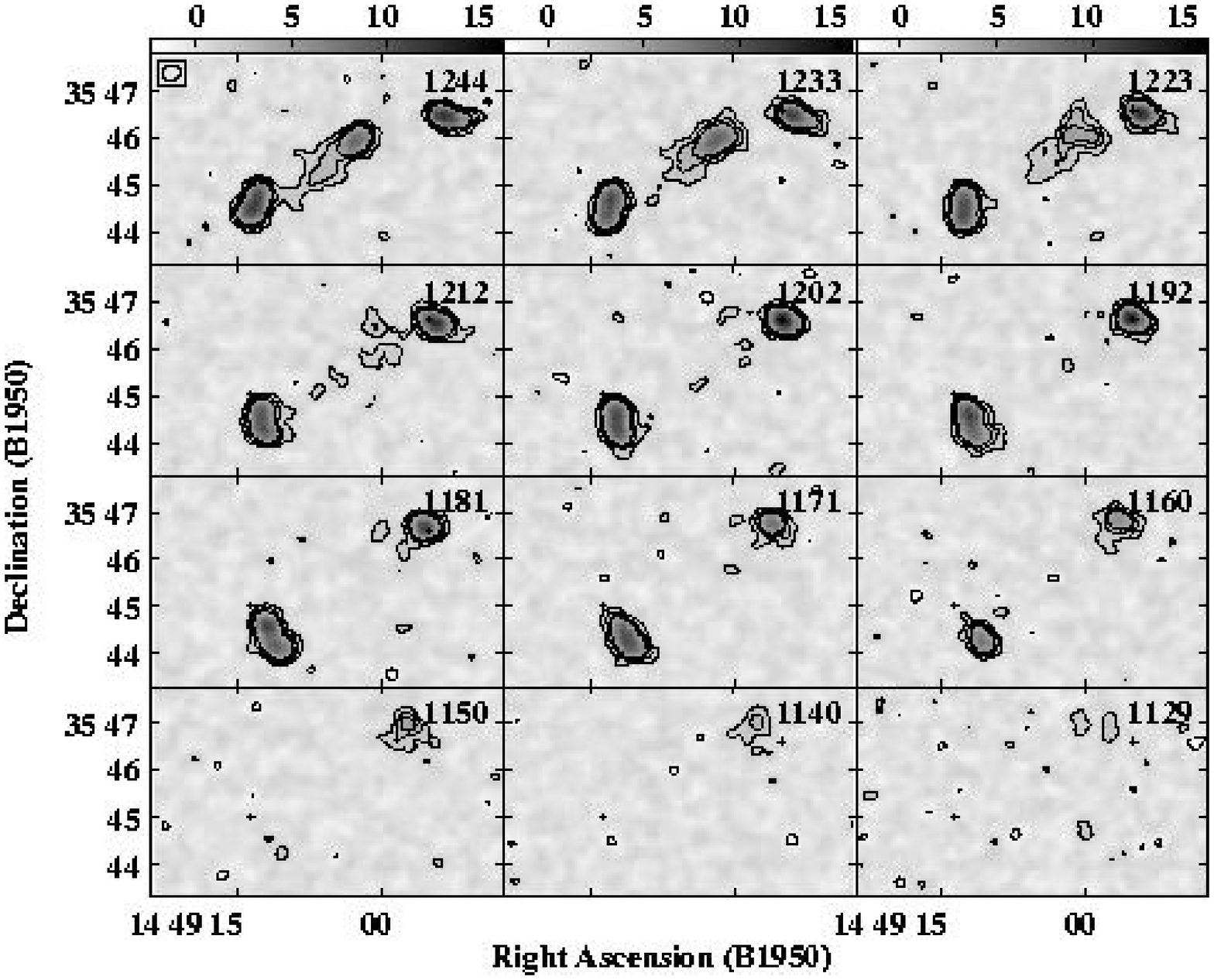}

\clearpage

\includegraphics[width=\linewidth]{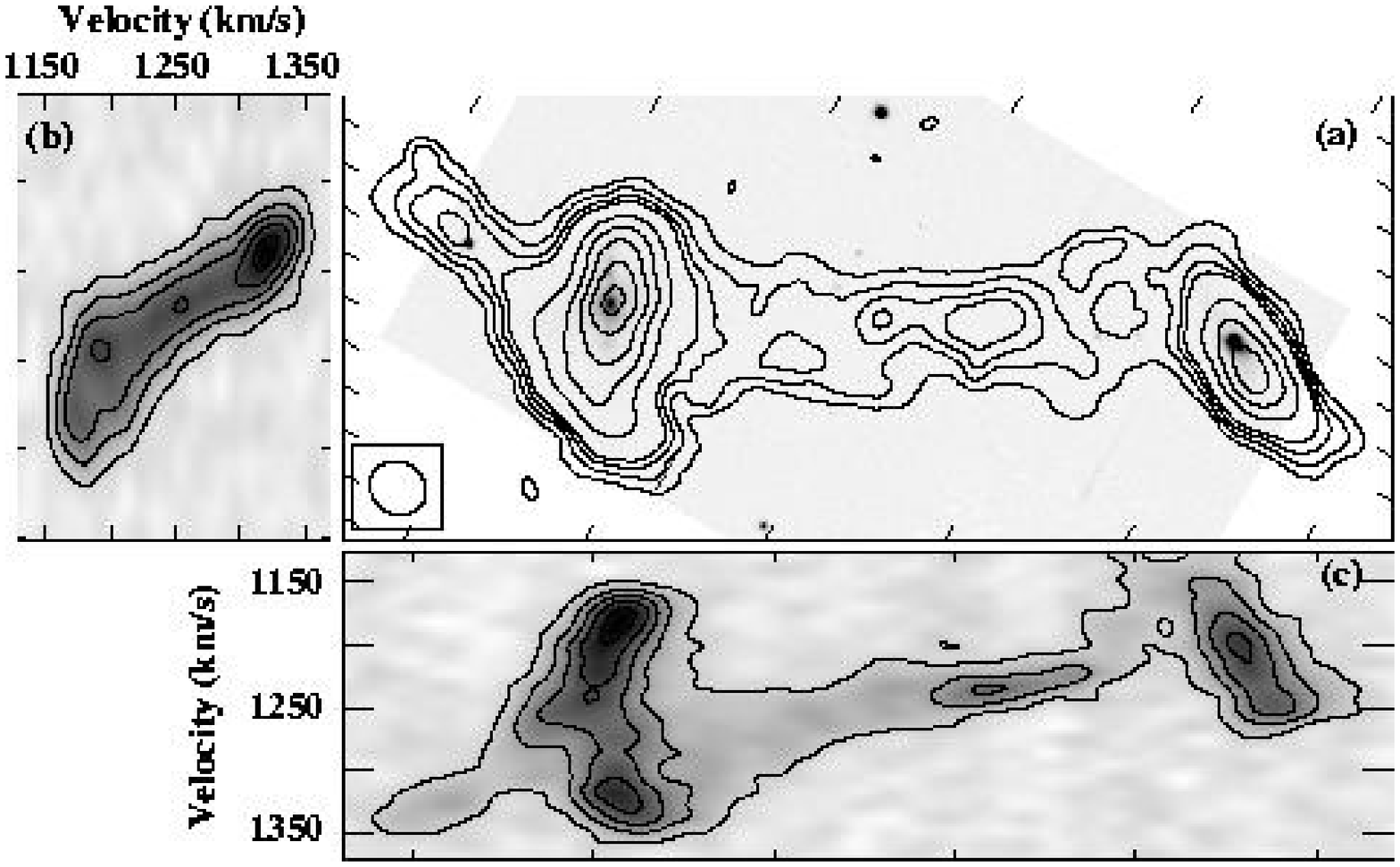}

\figcaption[II~Zw~71: Total HI overlaid on B-band image]{
%%MOMENT MAP & POS-VEL PLOTS
  {\bf (a)}
    Total-\HI\ contours superposed on the optical (B band) image of
    Figure 1.  Shown are II~Zw~71 (east), II~Zw~70 (west), and the
    \HI\ cloud between the two objects.  The image has been rotated
    30\arcdeg\ so that the apparent major axis of the gaseous cloud
    is horizontal. Notice the extension of the streamer
    on the eastern side of II~Zw~71 (the unresolved optical source at
    that location is a star).  The 21\,cm
    synthesized beam is shown in the lower left corner.
  {\bf (b)} 
    Position-velocity cut along the apparent major axis of the 
    optical polar ring.  This cut is a single 5\arcsec\ pixel wide.
  {\bf (c)} 
    Position-velocity sum along the apparent major axis of the gaseous
    cloud.  The sum includes all 21\,cm line emission detected from
    the streamer  (see Section 2.2).
\label{mom0}}

\clearpage

\includegraphics[width=\linewidth]{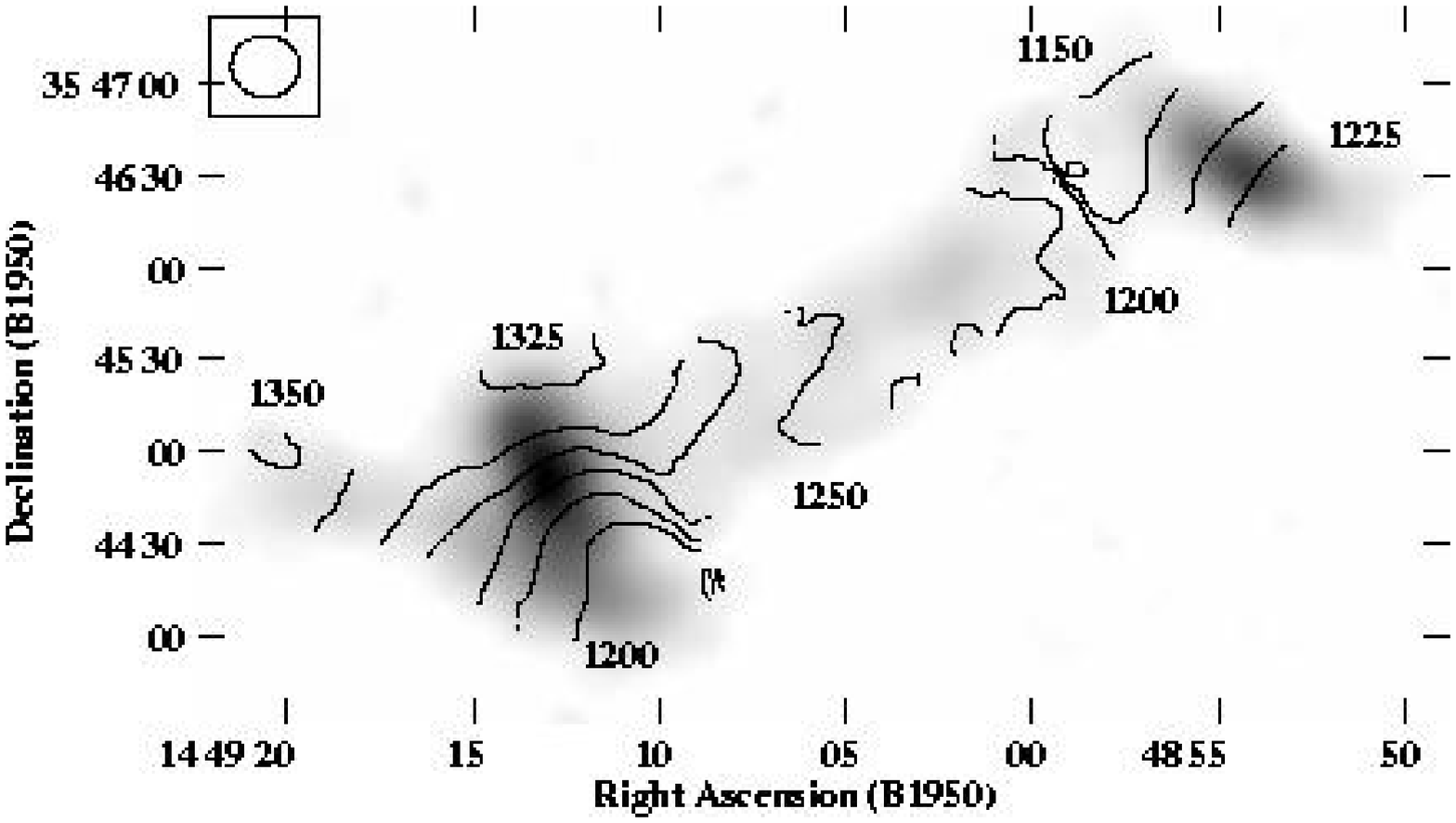}

\figcaption[II~Zw~71: Velocity field at 21\,cm]{
%%VELOCITY FIELD, OVERLAID ON TOTAL-HI MAP
  Intensity-weighted velocity field (contours) of the \zwab\ system,
  overlaid on the total-\HI\ map (greyscale) of Figure~3.  Contour 
  labels are in \kms, and all contours are separated by 25~\kms.  
  The 21\,cm synthesized beam is shown in the upper left corner.
\label{mom1}}

\clearpage

\includegraphics[width=\linewidth]{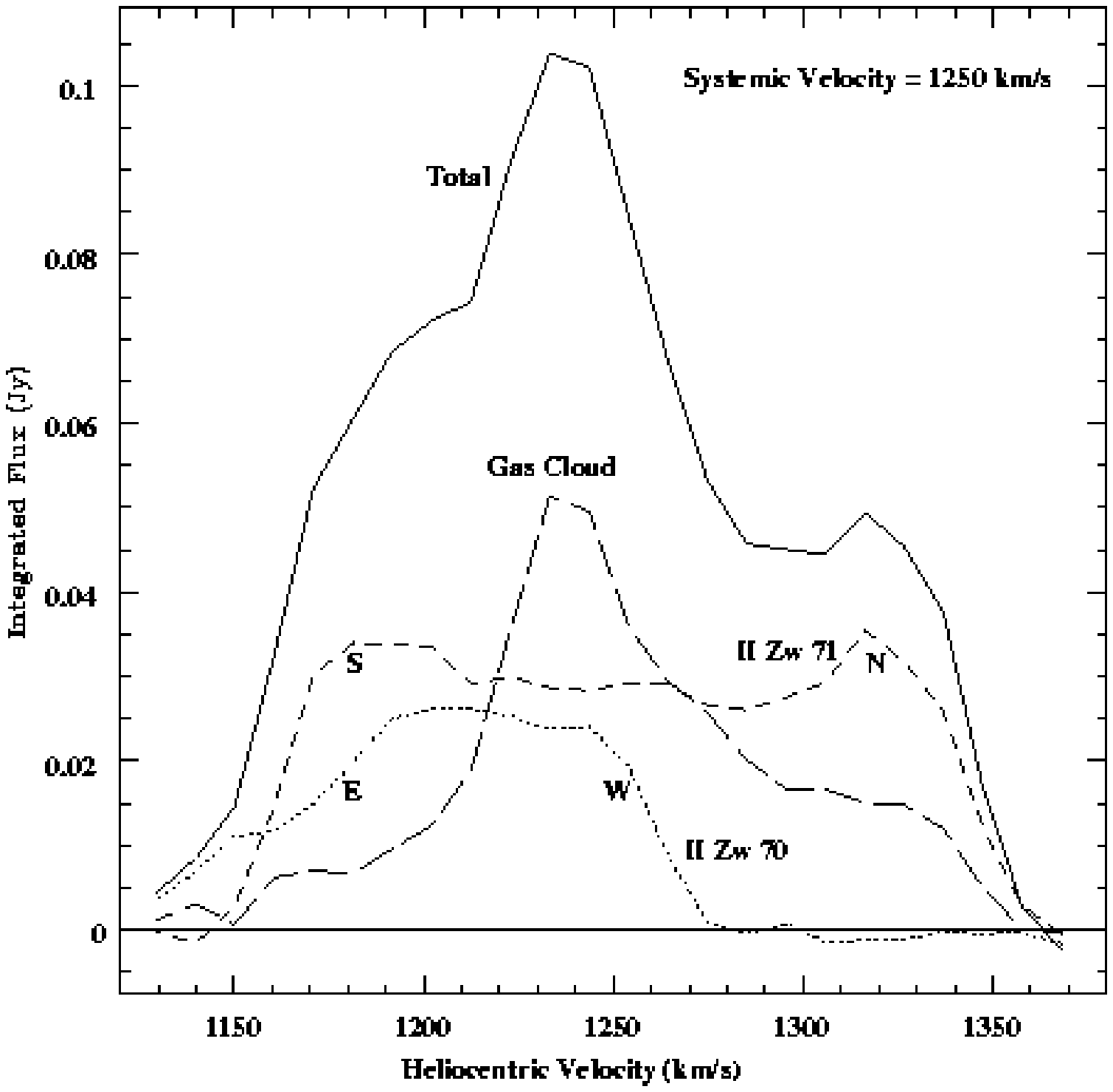}

\figcaption[II~Zw~71: 21\,cm spectrum]{
%%21cm INTEGRATED SPECTRUM
  21\,cm line profile for the II~Zw~71/70 system.  The solid line
  includes emission from the entire system, and the dashed lines
  show emission from each component, as labeled.  The
  high-velocity and low-velocity ends of the line profiles for \zwa\
  and \zwb\ are labeled to make it clear which part of each galaxy
  corresponds to these velocities.  The velocity width of the entire
  system at 20\%\ power is $\simeq$\,300~\kms, and the redshift of the
  system, determined from the midpoint of this range, is 1250~\kms.
\label{spectrum}}

\clearpage

\includegraphics[width=\linewidth]{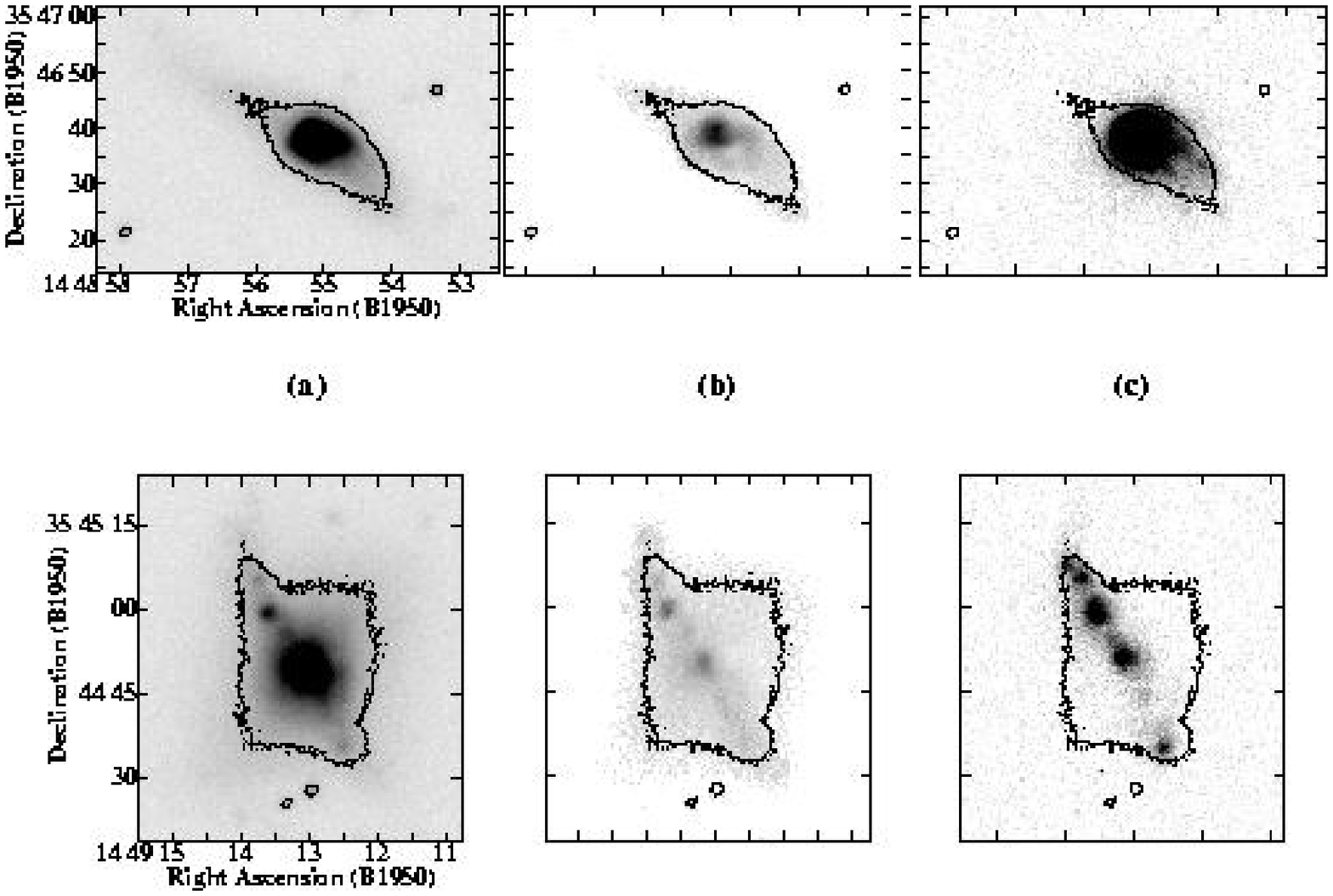}

\figcaption[Optical images]{
  Optical images of \zwa\ (top) and \zwb\ (bottom).  
  All images are to the same spatial scale.
  {\bf (a)}
    R band images. 
  {\bf (b)}
    B-R color maps of \zwa\ (top) and \zwb\ (bottom).  
    The contour represents the lowest R band contour from (a).
    Darker regions are bluer.
  {\bf (c)}
    \Halpha\ images.
    The contour represents the lowest R band contour from (a).
\label{BVimages}}

\begin{deluxetable}{lcc}
\tablenum{1}
\tablecaption{Instrumental Parameters: 21cm Observations of \zwa}
\tablehead{Parameter & \multicolumn{2}{c}{\zwa}}
\startdata
Array configuration 
   & C & D \\
Observing dates 
   & 02 Dec 1994 & 02 May 1995 \\
Time on-source (hr) 
   & 5.72 & 3.22 \\
Number of antennas\tablenotemark{a} 
   & 26 & 26 \\
Shortest, longest baseline (m) 
   & 73, 3400 & 35, 1030 \\
\cline{2-3}
Field center (1950)\tablenotemark{b}
   & \multicolumn{2}{c}{\phs14\hour 49\min 03\sec} \\
   & \multicolumn{2}{c}{$+$35\deg 45\arcmin 42\arcsec} \\
Velocity of band center, heliocentric (\kms)\tablenotemark{c} 
   & \multicolumn{1}{r}{\bf IF 1} & \multicolumn{1}{l}{1268} \\
   & \multicolumn{1}{r}{\bf IF 2} & \multicolumn{1}{l}{1232} \\
Number of velocity channels\tablenotemark{d} 
   & \multicolumn{2}{c}{128} \\
Frequency channel spacing (kHz)\tablenotemark{d} 
   & \multicolumn{2}{c}{12.2} \\
Velocity channel spacing (\kms)\tablenotemark{d} 
   & \multicolumn{2}{c}{2.6} \\
Final velocity of band center, (\kms)\tablenotemark{d}
   & \multicolumn{2}{c}{1247} \\
Final number of velocity channels\tablenotemark{d}
   & \multicolumn{2}{c}{49} \\
Final velocity channel spacing, (\kms)\tablenotemark{e}
   & \multicolumn{2}{c}{5.2} \\
FWHP of synthesized beam(\arcsec)\tablenotemark{d}
   & \multicolumn{2}{c}{22.5$\times$20} \\
RMS noise in channel maps (\mjyb)\tablenotemark{d}
   & \multicolumn{2}{c}{0.4} \\
\enddata
\tablenotetext{a}{
   Only 26 of the VLA's 27 antennas were used for these observations,
   as one antenna was allocated for VLBI observations.  
   }
\tablenotetext{b}{
   This parameter was the same for all observations of this galaxy.
   }
\tablenotetext{c}{
   This galaxy was observed with two offset bandpasses (IF's). 
   All parameters other than the central velocity were
   identical for these IF's.  The final maps were produced by combining
   data from both bandpasses.
   }
\tablenotetext{d}{
   The numbers quoted here are for the final maps after combining the
   data from both arrays and both IF's.  The edges of the two
   bandpasses were cropped to eliminate the noisy end-channels, and the
   remaining data were combined to make a single dataset with a wide
   enough bandwidth to contain the velocity range of both galaxies.
   }
\tablenotetext{e}{
   The final maps were smoothed to 5.2~\kms\ velocity resolution.  The
   high velocity resolution of the original observations was necessary 
   for an independent project by E. Brinks involving the same galaxy
   pair.
   }
\end{deluxetable}
\clearpage

\begin{deluxetable}{lcccc}
\tablenum{2}
\tablecaption{Optical magnitudes \&\ colors in the \zwab\ system}
\tablehead{
& & & \multicolumn{2}{c}{Component} \\
\cline{4-5}
Quantity\tablenotemark{a} & \zwa& \zwb & host galaxy & polar ring \\
}
\startdata
%
%    IIZw70	IIZw71	IIZw71(host)	IIZw71(ring)
%
m$_B$
   & 14.78 & 14.42 & $\simeq$\,14.7\tablenotemark{b} & \nodata \\
m$_V$
   & 14.45 & 13.99 & \nodata & \nodata \\
m$_R$
   & 14.23 & 13.59 & \nodata & \nodata \\
B-V
   &  &  & 0.56 & 0.36 \\
B-R
   &  &  & 0.99 & 0.65 \\
V-R
   &  &  & 0.43 & 0.30 \\
M$_B$
%   & -15.9 & -16.3 & $\simeq$\,-16.0 & \nodata \\  Ho=100
   & -16.5 & -16.9 & $\simeq$\,-16.6 & \nodata \\  %Ho=75
L$_B$ ($L_\odot$)
%   & 3.6\,$\times$\,$10^{8}$ 
%   & 5.0\,$\times$\,$10^{8}$ 
%   & $\simeq$\,3.9\,$\times$\,$10^{8}$ & \nodata \\
   & 6.2\,$\times$\,$10^{8}$ 
   & 8.7\,$\times$\,$10^{8}$ 
   & $\simeq$\,6.7\,$\times$\,$10^{8}$ & \nodata \\
\enddata
\tablenotetext{a}{
   No extinction corrections were made to the optical magnitudes to
   account for the orientation of the galaxies.
   }
\tablenotetext{b}{
   An approximate B magnitude for the host galaxy only was calculated by
   subtracting off the brightest parts of the polar ring from the B band
   image using a scaled version of the H-alpha line image.
   }
\end{deluxetable}

\begin{deluxetable}{lrccc}
\tablenum{3}
\tablecaption{
   Physical Parameters for the \zwab\ system
   }
\tablehead{
   & & \multicolumn{3}{c}{Source Name}\\
   \cline{3-5}
   \multicolumn{2}{l}{{\sc Parameter}} & \zwa & \zwb & streamer
   }
\startdata
%
%    II Zw 70   II Zw 71   streamer
%
\multicolumn{2}{l}{UGC number}
   & UGC~9560 & UGC~9562 & --- \\
Position (B1950) & $\alpha\,=$
   & \phs14\hour49\min13\sec & \phs14\hour48\min55\fs1
   & \\
   & $\delta\,=$
   & $+$35\deg44\arcmin47\arcsec & $+$35\deg46\arcmin37\arcsec
   & \\
\multicolumn{2}{l}{21\,cm Redshift (\kms)\tablenotemark{a}}
   & 1195 & 1250 & 1200 - 1345 \\
\multicolumn{2}{l}{Rotation speed (\kms)\tablenotemark{b}}
   & 67 & 95 & \nodata \\
\multicolumn{2}{l}{Angular diameter at 21cm}
   & $\simeq$\,90\arcsec & $\simeq$\,100\arcsec & \nodata \\
\multicolumn{2}{l}{\HI\ line integral (Jy~\kms)\tablenotemark{c}}
   & 2.6 & 6.2 & 3.2 \\
\multicolumn{2}{l}{Systemic velocity (\kms)\tablenotemark{d}}
   & 1310 & 1366 & \nodata \\
\multicolumn{2}{l}{Distance ($h^{-1}$~Mpc)\tablenotemark{e}}
   & 13.6 & 13.6 & 13.6 \\
\multicolumn{2}{l}{Linear radius in \HI\ ($h^{-1}$~kpc)}
   & 3.0 & 3.3 & \nodata \\
\multicolumn{2}{l}{\Mhi\ ($h^{-2}$~\Msun)\tablenotemark{c}}
   & $1.0$\,$\times$\,$10^8$ 
   & $\simeq$\,$2.7$\,$\times$\,$10^8$ 
   & $\simeq$\,$1.4$\,$\times$\,$10^8$ \\
\multicolumn{2}{l}{Total Mass, M$_{dyn}$ 
   ($h^{-1}$~\Msun)}
   & $6.2$\,$\times$\,$10^{9}$ 
   & $14$\,$\times$\,$10^{9}$ 
   & \nodata \\
\multicolumn{2}{l}{M$_{\rm HI}$/M$_{dyn}$
   ($h^{-1}$)}
   & 0.02 & 0.02 & \nodata \\
\multicolumn{2}{l}{M$_{\rm HI}$/L$_B$}
   & 0.32 & 0.55\tablenotemark{f} & \nodata \\
\enddata
\tablenotetext{a}{
   Midpoint between the 20\%\ level of the 21\,cm profile.
   }
\tablenotetext{b}{
   Half-width of the 21\,cm line at 20\%\ power
   }
\tablenotetext{c}{
   The 21cm line integrals and \HI\ masses listed here are based upon
   the line profiles shown in Figure~5.  The method used to estimate the
   amount of emission associated with each object is described in
   Section~2.3.
   }
\tablenotetext{d}{
   After correcting for solar motion of 300~\kms\ towards $l\,=\,90$\deg,
   $b\,=\,0$\deg.
   }
\tablenotetext{e}{
   All distance-related parameters are written in terms of the
   dimensionless unit $h$\,$=$\,\Ho/$100$~\kms~Mpc.  The systemic velocity
   assumed for the system is 1364~\kms, and the same distance is used
   for all three components of the \zwa/\zwb/streamer system.  (The
   systemic velocity is based upon a redshift of 1249~\kms, which is
   the midpoint at 20\%\ power of the global line profile in Figure~5.)
   }
\end{deluxetable}

\end{document}